\documentclass{PoS}

\usepackage{multirow}
\def\f2r{\mathrm{Ref_{2R}}}
\def\rr{\mathrm{Re\rho_{2}}}
\def\ri{\mathrm{Im\rho_{3}}}
\def\onebmw2{\frac{1}{m_W^2}}

\title{Polarization of the top quark as a probe of its chromomagnetic and chromoelectric 
couplings in single-top production  at the Large Hadron Collider}

\ShortTitle{Top Asymmetries and anomalous top couplings}

\author{\speaker{Saurabh D. Rindani}\\
        Theoretical Physics Division, Physical Research Laboratory,\\
        Navrangpura, Ahmedabad 380 009, India,\\
        E-mail: \email{saurabh@prl.res.in}}

\author{Pankaj Sharma\\
        Center of Excellence in Particle Physics (CoEPP),\\
        The University of Adelaide, Adelaide, Australia\\
        E-mail: \email{pankaj.sharma@adelaide.edu.au}}

\author{Anthony W. Thomas\\
        Center of Excellence in Particle Physics (CoEPP),\\
        The University of Adelaide, Adelaide, Australia\\
        E-mail: \email{anthony.thomas@adelaide.edu.au}}

\abstract{We study the sensitivity of the Large Hadron Collider (LHC) to top quark chromomagnetic (CMDM) 
and chromoelectric (CEDM) dipole moments and $Wtb$ effective couplings in single-top production in
association with a $W^-$ boson, followed by semileptonic decay of the top. We calculate the top
polarization and the effects of these anomalous couplings on it at two centre-of-mass (cm)
energies, 8 TeV and 14 TeV. As a measure of top polarization, we look at decay-lepton angular
distributions in the laboratory frame,  without requiring reconstruction of the rest frame of the
top, and study the effect of the anomalous couplings on these distributions. We construct certain
asymmetries to study the sensitivity of these distributions to top-quark couplings. The $Wt$
single-top production mode helps to isolate the anomalous $ttg$ and $Wtb$ couplings, in contrast to
top-pair production and other single-top production modes, where other new-physics effects can
also contribute. We determine individual limits on the dominant couplings, viz., the real part of
the CMDM $\rr$, the imaginary part of the CEDM $\ri$, and the real part of the tensor $Wtb$
coupling $\f2r$, which may be obtained by utilizing these asymmetries at the LHC. We also obtain
simultaneous limits on pairs of these couplings taking two couplings to be non-zero at a time.
}

\FullConference{8th International Workshop on Top Quark Physics, TOP2015\\
		14-18 September, 2015\\
		Ischia, Italy}

\begin{document}

\section{Introduction}
The top quark is the heaviest fundamental particle discovered so far, with mass $m_t=173.2\pm 0.9$ GeV.
 Because of its large mass, the top-quark life time is very short and it decays spontaneously
before any non-perturbative QCD effects can force it into a bound state. So by
studying the kinematical distributions of top decay products, it is, in principle, possible to
measure the top polarization in any top production process. 

Top quark couplings to a gluon can be defined in a general way as
\begin{equation}
\Gamma^\mu=\rho_1\gamma^\mu +\frac{2 i}{m_t}\sigma^{\mu\nu}\left(\rho_2+i\rho_3\gamma_5\right)
q_\nu,
\end{equation}
where $\rho_2$ and $\rho_3$ are top quark CMDM and CEDM form factors.
Of these, the $\rho_2$ term is CP even, whereas the $\rho_3$ term is CP odd. In the
SM, both $\rho_2$ and $\rho_3$ are zero at tree level.

In this work, we study $Wt$ production at the LHC in the presence of 
top CMDM and CEDM. In particular, we examine the possibility of using 
top polarization and angular observables, constructed from top decay products in the
laboratory frame, to measure these couplings. 
We construct an asymmetry out of the azimuthal distribution of charged leptons in the lab frame and show that 
it has all the characteristics of top polarization. 
More details about top polarization and azimuthal asymmetry can be found in \cite{delDuca:2015gca} and references therein. 
This asymmetry was first defined in \cite{Godbole:2010kr}.
Subsequently, it has been studied extensively in the context of constraining top chromo-dipole couplings in top pair production 
\cite{Biswal:2012dr}, in $tW$ production \cite{Rindani:2015vya}; and anomalous $Wtb$ couplings \cite{Rindani:2011pk} and 
CP violation \cite{Rindani:2011gt} in single-top production at the LHC. In the context of two Higgs doublet models (2HDM), it has been 
used to determine $\tan\beta$ \cite{Huitu:2010ad, Godbole:2011vw} and distinguish different 2HDMs \cite{Rindani:2013mqa}.

\section{Top polarization and angular asymmetries of charged lepton}
Top polarization can be determined through the angular distribution of its decay products. 
The angular distribution of a decay product $f$ for a top-quark ensemble has the form 
 \begin{equation}
 \frac{1}{\Gamma_f}\frac{\textrm{d}\Gamma_f}{\textrm{d} \cos \theta _f}=\frac{1}{2}(1+\kappa _f P_t \cos \theta _f).
 \label{topdecaywidth}
 \end{equation}
Here $\theta_f$ is the angle between the momentum of fermion $f$ and the top spin vector in the
top rest frame. In this work, we study the lepton azimuthal distribution, in the lab frame, as a qualitative measure of top polarization. 
In the lab frame, we define the lepton azimuthal angle with respect to the top-production plane chosen as the $x$-$z$ plane,
with the convention that the $x$ component of the top momentum is positive. 

\begin{figure}[h!]
\begin{center}
\includegraphics[scale=0.45]{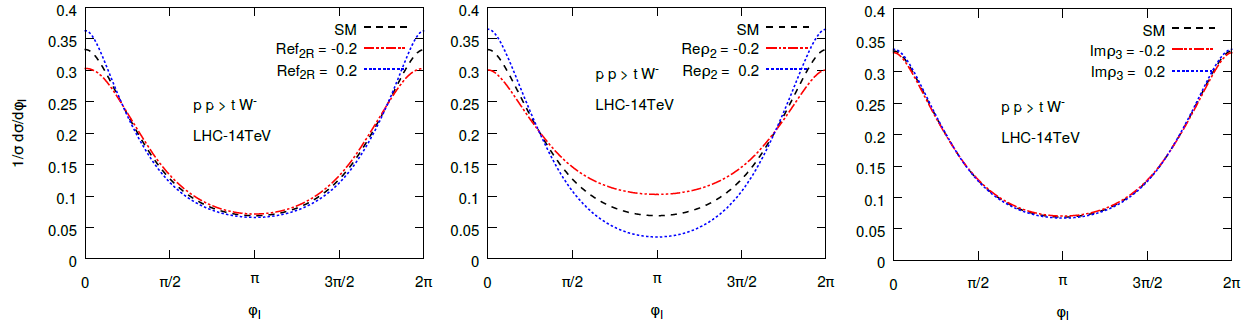}
\caption{ The normalized azimuthal distribution of the charged lepton in associated-$W$t
single-top production at the LHC14 for anomalous couplings $\f2r$, $\rr$ and $\ri$.} 
\label{dist-azi}
\end{center}
\end{figure}

In Fig. \ref{dist-azi}, we show the lepton azimuthal distribution for different anomalous top couplings at the 14 TeV LHC. 
We find that these distributions show the best sensitivity to $\rr$ and $\f2r$. To quantify the sensitivity of these  
distributions, we further define an asymmetry in terms of partially integrated cross sections
\begin{equation}
 A_{\phi}=\frac{\sigma(\cos \phi_\ell >0)-\sigma(\cos
\phi_\ell<0)}{\sigma(\cos \phi_\ell >0)+\sigma(\cos \phi_\ell<0)},
\label{aziasy}
\end{equation}

\begin{figure}[h!]
\begin{center}
\includegraphics[scale=0.38]{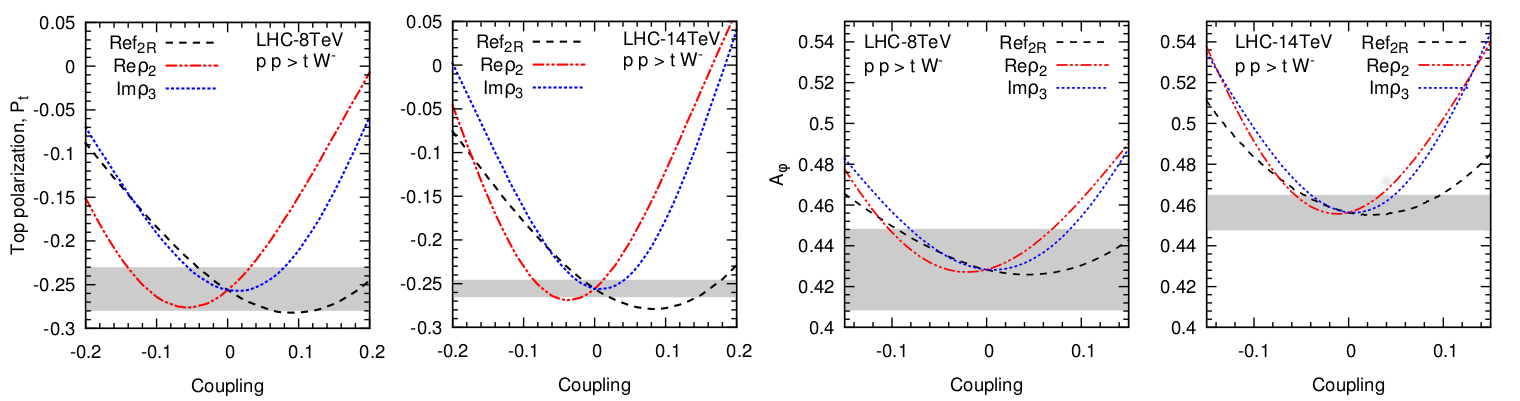} 
\end{center}
\caption{ The top polarization and azimuthal asymmetry $A_\phi$ for $tW^-$ production as a function of top-anomalous couplings. } 
\label{polcoup-full}
\end{figure}

In Fig. \ref{polcoup-full}, we display the top polarization and azimuthal asymmetry for different anomalous top couplings 
at the 8 and 14 TeV LHC. We also study the sensitivity of the observables to the measurement of the anomalous couplings.
The 1$\sigma$ limits on $\f2r$, $\rr$ and $\ri$ are given in Table \ref{lim} for LHC8 and LHC14, where we 
assume only one anomalous coupling to be non-zero at a time. We have assumed measurements
on a $tW^-$ final state and only one leptonic channel. Including the conjugate process and other leptonic 
decays of the top would improve the limits further.

\begin{figure}[h!]
\begin{center}
\includegraphics[scale=0.5]{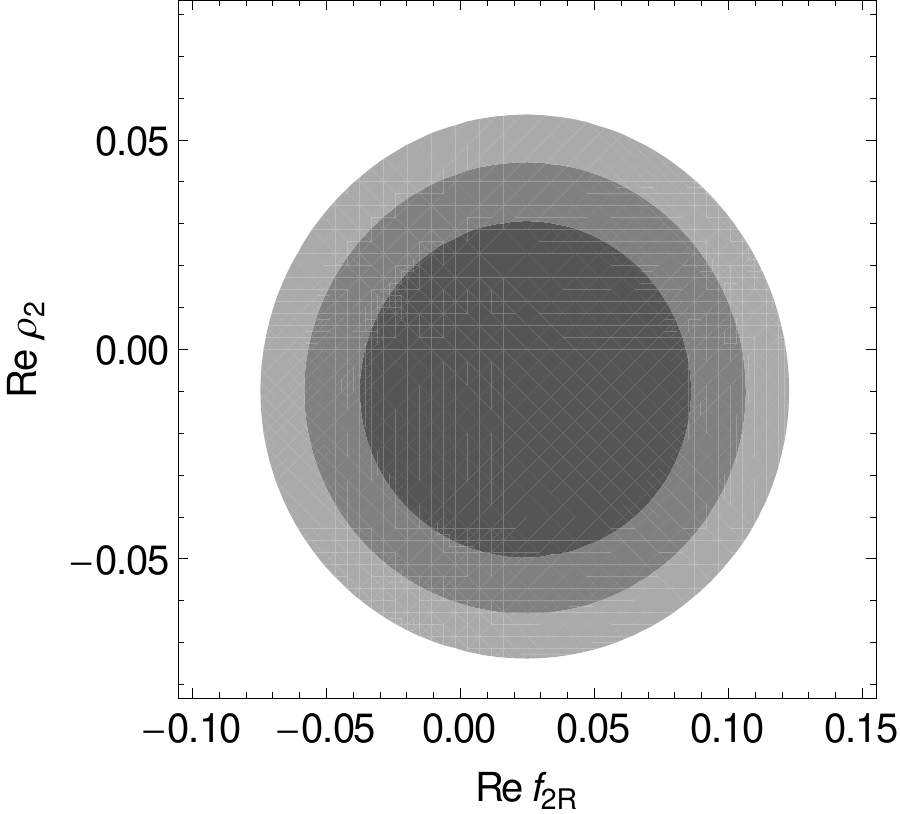} 
\includegraphics[scale=0.5]{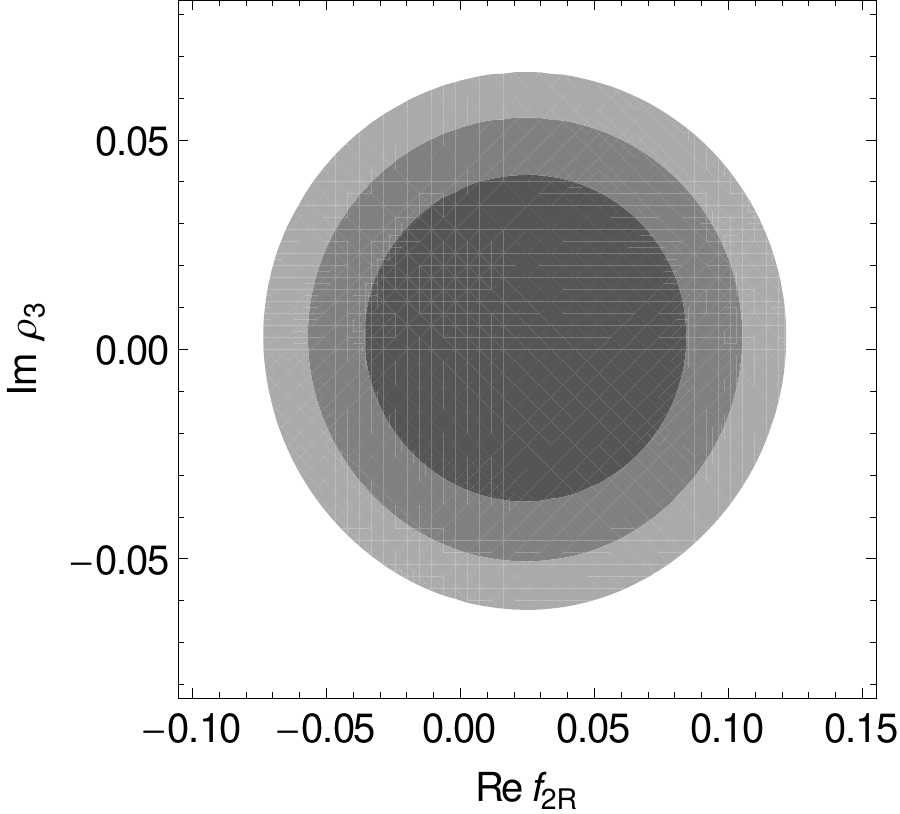} 
\includegraphics[scale=0.5]{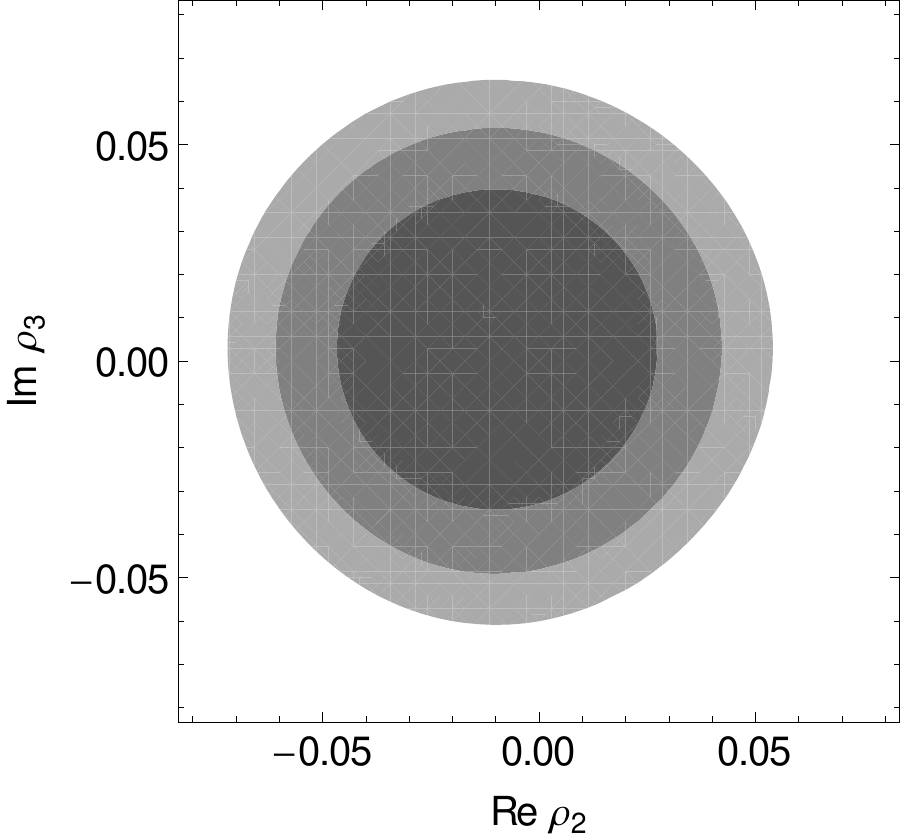} 
\caption{The 1$\sigma$ (central), 2$\sigma$ (middle) and 3$\sigma$ (outer)
 CL regions in the $\f2r$-$\rr$ plane (left), $\f2r$-$\ri$ plane (center) and 
$\rr$-$\ri$ plane (right) allowed by the measurement of $A_{\phi}$ at the 
LHC14. } 
\label{lim-all}
\end{center}
\end{figure}

 Apart from the 1$\sigma$ limits shown in Table \ref{lim-all}
there are other disjoint intervals which could be ruled out if no deviation from the SM were
observed for $P_t$ and $A_{\phi}$. This is apparent from Fig. \ref{polcoup-full}. The additional
allowed intervals for $\f2r$ and $\rr$ from $P_t$ measurement are [0.158, 0.205] and [-0.80,
-0.65] for LHC14, respectively\footnote{[a, b] denotes the allowed values of the coupling $f$ at
the 1$\sigma$ level, satisfying $a<f<b$.}. It is seen that the top polarization, $P_t$, and
azimuthal asymmetry, $A_{\phi}$, of the charged lepton are more sensitive to negative values of
the anomalous couplings $\f2r$ and positive values of $\rr$. 

\begin{table}[h!]
\begin{center}
\begin{tabular}{|c|c|c|c|c|}
\hline
\hline
			& Observable & 	$\f2r$	&  $\rr$ & 	$\ri$	\\
 \hline
 \multirow{2}{*}{8 TeV}  &  $P_t$ 			&[$-0.030$, $0.032$]	& [$-0.028$, $0.019$] 	&[$-0.038$, $0.065$]\\
 						 &  $A_\phi$		&[$-0.060$, $0.140$]	& [$-0.080$, $0.050$] 	&[$-0.055$, $0.070$]\\
\hline
 \hline
 \multirow{2}{*}{14 TeV} & $P_t$ 			&[$-0.010$, $0.010$]	& [$-0.009$, $0.009$] 	&[$-0.020$, $0.035$]\\
 						 & $A_\phi$ 		&[$-0.031$, $0.081$]	& [$-0.045$, $0.020$] 	&[$-0.030$, $0.040$]\\
\hline
\hline
  \end{tabular}
\caption{Individual limits on anomalous couplings $\f2r$, $\rr$ and $\ri$ which may be 
obtained by the measurement of the observables at 8 and 14 TeV with integrated luminosities of 20 and 30 fb$^{-1}$ 
respectively.}
 \label{lim}
\end{center}
 \end{table}

We also obtain simultaneous limits (taking two couplings out of $\f2r$, $\rr$ and $\ri$ 
non-zero simultaneously) on these anomalous couplings that may be obtained by the 
measurements of asymmetries.
For this, we perform a $\chi^2$ analysis to fit all the observables to within $f\sigma$ of
statistical errors in the measurement of the observable.
In Fig. \ref{lim-all}, we show the 1$\sigma$, 2$\sigma$ and 3$\sigma$
regions in $\f2r-\rr$ plane, $\f2r-\ri$ plane and $\rr-\ri$ plane allowed by the measurement of
the asymmetry $A_\phi$. From the plots shown in Fig. \ref{lim-all}, we find that the strongest
simultaneous limits are [$-0.03$, 0.08] on $\f2r$, [$-0.05$, 0.02] on $\rr$  and [$-0.03$, 0.03]
on $\ri$, at the $1\sigma$ level.

\section{Conclusions}\label{conclusions}

In conclusion, we have shown that top polarization, and subsequent decay-lepton distributions can
be used to obtain fairly stringent limits on chromomagnetic and chromoelectric top couplings from
the existing 8 TeV run of the LHC. The limits could be improved by the future runs of the LHC at
14 TeV.


\begin{thebibliography}{99}
\bibitem{delDuca:2015gca} 
  V.~del Duca and E.~Laenen,
  arXiv:1510.06690 [hep-ph].
  
\bibitem{Godbole:2010kr} 
  R.~M.~Godbole, K.~Rao, S.~D.~Rindani and R.~K.~Singh,
  JHEP {\bf 1011}, 144 (2010).
  
\bibitem{Rindani:2015vya} 
  S.~D.~Rindani, P.~Sharma and A.~W.~Thomas,
  JHEP 10 (2015) 180. 

\bibitem{Biswal:2012dr} 
  S.~S.~Biswal, S.~D.~Rindani and P.~Sharma,
  Phys.\ Rev.\ D {\bf 88}, 074018 (2013).
  
\bibitem{Rindani:2011pk} 
  S.~D.~Rindani and P.~Sharma,
  JHEP {\bf 1111}, 082 (2011)  

\bibitem{Rindani:2011gt} 
  S.~D.~Rindani and P.~Sharma,
  Phys.\ Lett.\ B {\bf 712}, 413 (2012).
  
\bibitem{Huitu:2010ad} K.~Huitu, S.~Kumar Rai, K.~Rao, S.~D.~Rindani and P.~Sharma, 
 JHEP {\bf 1104}, 026 (2011).
 
\bibitem{Godbole:2011vw} 
  R.~M.~Godbole, L.~Hartgring, I.~Niessen and C.~D.~White,
  JHEP {\bf 1201}, 011 (2012) 
  
\bibitem{Rindani:2013mqa} S.~D.~Rindani, R.~Santos and P.~Sharma,
  JHEP {\bf 1311}, 188 (2013). 
      
\end{thebibliography}
\end{document}